\documentclass[letterpaper,prl,twocolumn,superscriptaddress]{revtex4}
\usepackage{graphicx,bm}
\usepackage{color}
\begin{document}
\newcommand{\mybm}[1]{\mbox{\boldmath$#1$}}
\newcommand{\mysw}[1]{\scriptscriptstyle #1}

\newcommand{\edit}[1]{\textcolor{red}{#1}}

\title{Quantized magnetic moment at the edge of a single-walled carbon nanotube}

\author{Horng-Tay Jeng}
\affiliation{Institute of Physics, Academia Sinica, Nankang, Taipei 11529, Taiwan}
\affiliation{Department of Physics, National Tsing-Hua University, Hsinchu 300, Taiwan}

\author{Masaki Oshikawa}
\affiliation{Institute for Solid State Physics, University of Tokyo, Kashiwa 227-8581 Japan}

\author{Hsiu-Hau Lin}
\affiliation{Department of Physics, National Tsing-Hua University, Hsinchu 300, Taiwan}
\date{\today}

\begin{abstract}
We investigate many-body effects near the edge of a single-walled carbon nanotube and find it turns magnetic with quantized edge moment solely depends on the chiral vector, i.e. the topology of the carbon nanotube.
The distribution of the edge moment reveals an approximate supersymmetry
even in a realistic model.
Our findings demonstrate crucial importance of the interplay between the electronic correlation and the edge morphology.
The edge moment
provides an excellent candidate for nanomagnet farbrication which
has potential applications in
biological and chemical detections, and in spintronics.
\end{abstract}
\maketitle

A single-walled carbon nanotube (SWNT)~\cite{Charlier07} can be viewed as rolling up the two-dimensional graphene~\cite{Novoselov04,Novoselov05,Zhang05,Geim07} into cylindrical shape with radius of the order of nanometers.
Its quasi-one-dimensionality leads to pronounced many-body effects, manifest in optical transitions for electron-hole excitons~\cite{Wang05,Dukovic05}, electronic bandgaps by scanning tunnelling spectroscopy~\cite{Lin10} etc. and leads to exotic ground states including Luttinger liquids~\cite{Bockrath99,Ishii03}, Wigner crystal~\cite{Deshpande08} and Mott-insulating spin liquids~\cite{Deshpande09}.

Because the open boundary in the honeycomb network changes the electronic states dramatically, it is interesting to explore how correlation effects reshape the physical properties. 
For the simplest zigzag edge, both analytic and numeric approaches~\cite{Fujita96,Nakada96} indicate additional zero-energy excitations at the edge, while for the armchair, none of these evanescent modes exist.
However, counting the total number of the zero modes for generic chiral SWNTs remains unknown.
Since these edge states are degenerate, inclusion of mutual interactions between electrons induce ferromagnetic coupling and promotes magnetism in SWNTs.
In fact, correlation-induced magnetic moment has be speculated for quite a while~\cite{Wakabayashi99,Okada01,Hikihara03,Son06} but its magnitude hasn't been quantified yet.

The central goal of this Letter is to show how edge moment emerges in generic SWNTs.
Cutting a single-walled carbon nanotube perpendicularly, we investigate the many-body effects near the hydrogenated edge and find the nanotube turns magnetic with quantized edge moment solely depends on the chiral vector, i.e. the topology of the carbon nanotube.
Meanwhile, the spin density for the edge moment only shows up in one of the sublattices for the honeycomb, revealing an approximate supersymmetry~\cite{Mou04,Lin05} even with the inclusion of the realistic band structures.
Our findings demonstrate that the interplay between the electronic correlation and the edge morphology is of crucial importance at nanoscale.
The edge moment, solely depending on the topology of the nanotube, provides an excellent candidate for nanomagnet farbrication which adds a spin flavor to biological~\cite{Heller06} and chemical detections~\cite{Snow05} and also potential applications in spintronics~\cite{Ando00,Huertas-Hernando06,Nowack07,Kuemmeth08} for carbon nanotubes.

\begin{figure}
\centering
\includegraphics[width=8cm]{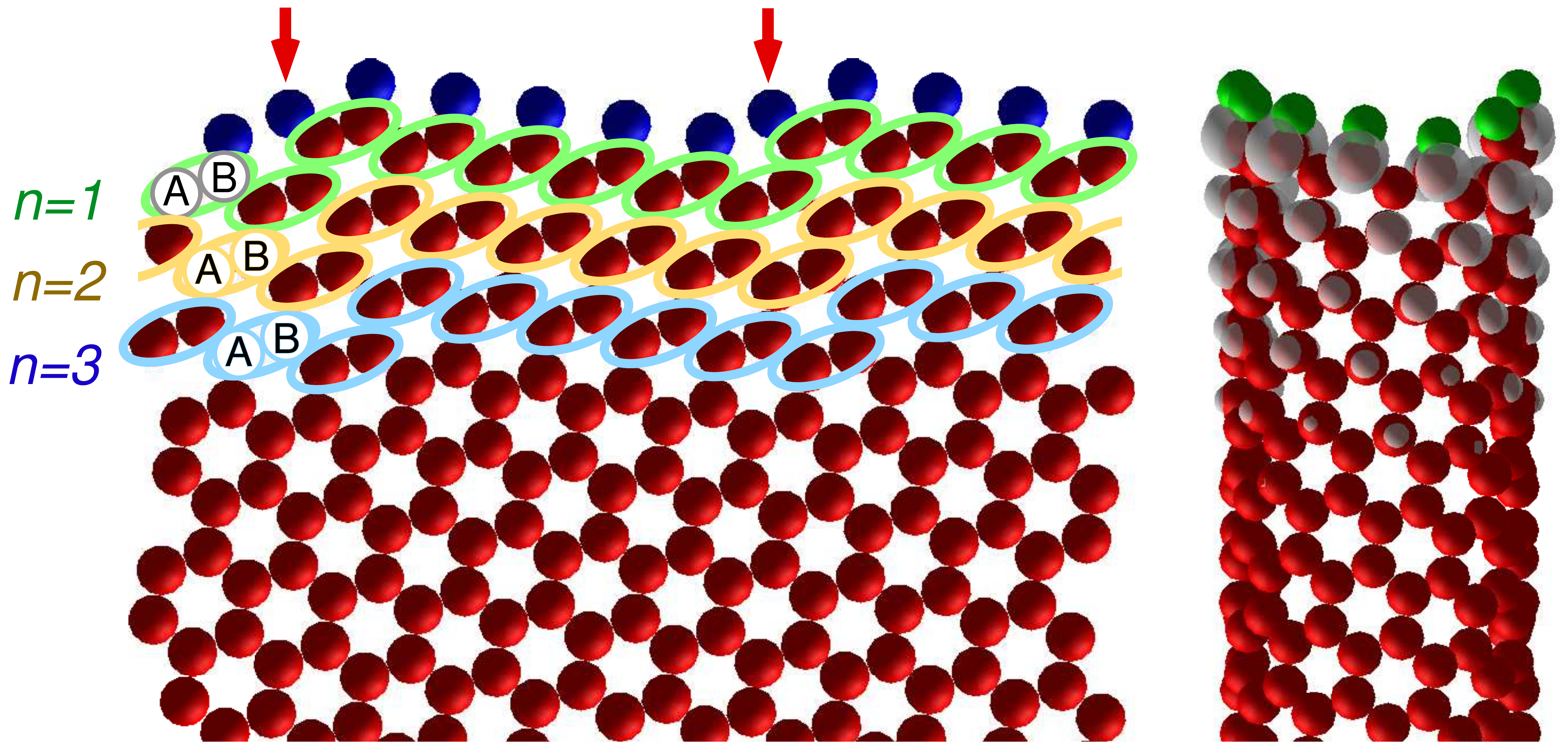}
\caption{\label{Fig1}
A SWNT and its flattened form with chiral vector $(10,2)$, where the red and green spheres denote C and H atoms respectively. The spin density (grey color) arisen from electron-electron interactions appears near the edge and decays rapidly into the bulk.
}
\end{figure}

Let us start with how the topology of a SWNT is classified.
The chiral vector, composed of two integers $(N,M)$, denotes a lattice vector in the honeycomb lattice of graphene and label the waist uniquely in the roll-up SWNT.  
To remove the redundancy from the hexagonal symmetry in the honeycomb structure, we fix the convention for the chiral vector, $N \geq M \geq 0$ ~\cite{Hamada92}.
For $M=0$, this family of SWNTs is named ``zigzag" while the other known ``armchair" family carries the chiral vector, $N=M$.
Otherwise, they are called chiral in general.
To classify all possible edges of semi-infinite SWNTs is more tricky.
For a given chiral vector, there are infinite ways to cut the nanotubes.
Therefore, we only consider the primary edge by cutting along the chiral vector, i.e. the perpendicular cut around the nanotube.
For simplicity, we assume the primary edge is terminated by hydrogen atoms.
As an example, a SWNT with chiral vector $(10,2)$ and its flattened form are shown in Fig. 1.
The honeycomb structure consists of two sublattices $A$ and $B$.
The primary edge is obtained by cutting the SWNT perpendicularly.
The unit cell contains 12 carbon atoms on each sublattice and the hydrogenated edge is connected to 2 carbon atoms on the $A$-sites and 10 on the $B$-sites.
We also show the spin density determined by our first-principles
calculation, which will be explained later in detail;
the spin density arisen from electron-electron interactions appears near the edge and decays rapidly into the bulk.
It is surprising that the spin density profile has nodes on the $A$-sites, hinting an approximate supersymmetry in SWNT, even with the inclusion of realistic band structures.

We start with the Hubbard model on the honeycomb lattice,
\begin{eqnarray}
H = \sum_{x\in A, y \in B} t_{xy} 
[c^{\dag}_{x\sigma}c^{}_{y\sigma}+ c^{\dag}_{y\sigma}c^{}_{x\sigma}]
+ \sum_{x} U n_{x\uparrow} n_{x\downarrow},
\end{eqnarray}
where the hopping amplitude $t_{xy}=-t$ for nearest neighbors and vanishes otherwise.
The electron-electron interaction is approximated by the on-site repulsion $U>0$.
Although this Hubbard Hamiltonian is a simplified version for the realistic SWNT, it is still too hard to be solved exactly. 

\begin{figure}
\centering
\includegraphics[width=5.5cm]{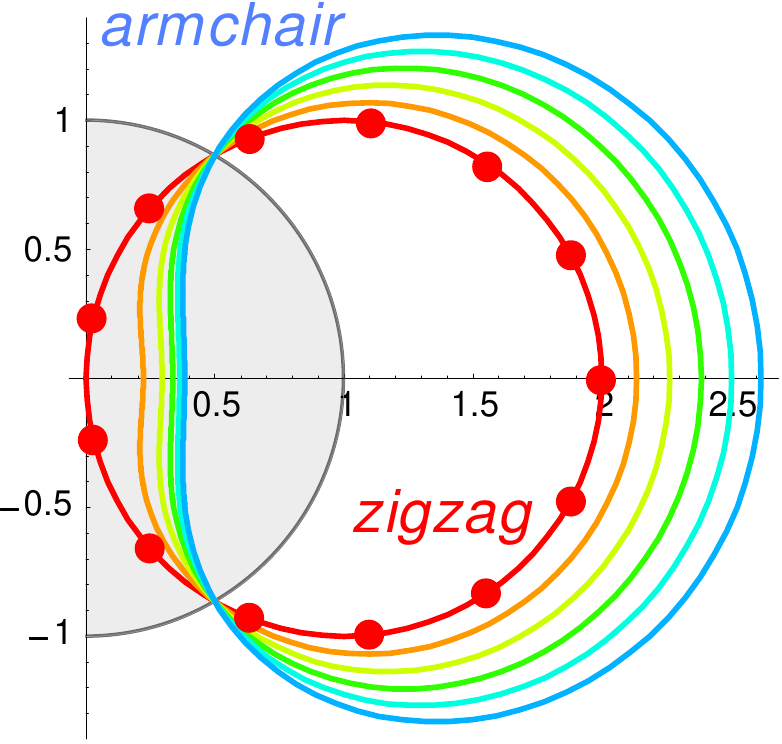}
\caption{\label{Fig2}
Graphical solutions for $z$ in the complex plane for SWNTs. Varying zigzag $C=0$ to armchair $C=1$, each family of SWNTs falls onto a universal curve and forms the set of the eye-ball contours. As an example, the red dots represent the non-unitary solutions $|z| \neq 1$ for semi-inifite $(13,0)$ zigzag SWNT. 
}
\end{figure}

Ignoring the on-site repulsion momentarily, the best strategy to pinpoint the edge states in the presence of the open boundary is by the hidden supersymmetry (SUSY)~\cite{Mou04,Lin05} in the honeycomb lattice.
The SUSY not only enforces these edge states at zero energy but also guarantees their wave functions vanish in one of the sublattices ($A$-sites in Fig. 1).
The non-vanishing wave function on sublattice $B$ can be solved by the generalized Bloch theorem~\cite{Lin05},
\begin{eqnarray}
\bm{\Psi}_e(n) = \sum_{i=1}^{N_{<}} (-1)^n c_{i} z_{i}^n \bm{\Phi}_{i},
\end{eqnarray}
where $\Phi_{i}$ is a $(N+M)$ dimensional column vector describing the all $B$-sites in the unit cell, $z$ is the eigenvalue of unit displacement operator and $n$ labels the number of the unit cells in the longitudinal direction.
The $(-1)^n$ factor is for notational convenience and $c_i$ are appropriate constants to satisfy the boundary conditions at the edge.
Since the translational invariance is lost in the longitudinal direction, in addition to the ($|z|=1$) plane-wave solutions, $N_{<}$ evanescent modes with $|z|<1$ appear.
As is clear in Fig. 1, there are $M/N$ constraints from $A/B$-sites at the edge.
However, the wave function has nodes on $A$-sites, rendering the $N$ constraints trivially satisfied.
Subtracting the lost degrees of freedom from the $M$ constraints, we end up with $N_e = N_{<}-M$ zero modes due to the edge of the SWNT.

The solutions of $z_i$ for the evanescent modes can be found as in the standard calculations for the band structures.
The characteristic equation from the generalized Bloch theorem turns out to be remarkably simple,
\begin{eqnarray}
\bigg(z-1\bigg)^{N} \left(\frac1z - 1\right)^{M} =1.
\label{z-eq}
\end{eqnarray}
Introducing the complex function $f(z) = (z-1)^N (1/z-1)^M$, the above equation is equivalent to $ \arg f(z) = 0$ (modulus $2\pi$) and $|f(z)|=1$, where the latter constraint gives a set of ``eye-ball" contours as shown in Fig. 2. Note that these contours only depend on the chirality $C \equiv M/N$, varying from zero (zigzag) to one (armchair). All contours intersect at two points at $z_{\pm}=e^{\pm i\pi/3}$, which correspond to the pair of the Dirac conducting channels.

All evanescent modes requires $|z|<1$ and thus must occur on the shorter segment between $z_{\pm}=e^{\pm i\pi/3}$. Furthermore, one can show that $\arg f(z)$ changes monotonically along the eye-ball contour. As a result, counting edge states is equivalent to tracking the winding number of $\arg f(z)$ passing through zero when going from $z_+$ to $z_-$. The factor $(z-1)^N$ in the complex function gives phase change $2\pi N /3$ while $(1/z-1)^M = (z-1)^M (-z)^{-M}$ gives $2\pi M/3 - (-2\pi M/3) = 4\pi M/3$. Therefore, the total phase change is $\Delta [\arg f(z)] = 2\pi (N+2M)/3$. For metallic SWNTs where $N-M$ is an integer multiple of three, straightforward analysis gives the winding number from the total phase change is $N_{<}=(N+2M)/3 -1$. Similarly, for semiconducting ones, the winding number is $N_< = [(N+2M+1)/3]$, where $[x]$ is the Gauss symbol taking the greatest integer less than or equal to $x$. Subtracting the number of constraints $M$,
\begin{eqnarray}
N_e &=&
\frac{N-M}{3}-1\quad  (\mbox{metallic}), 
\\
N_e &=& \left[\frac{N-M+1}{3}\right] \quad (\mbox{semiconducting}).
\end{eqnarray}

To investigate the correlation effects, let us consider the semiconducting SWNTs in weak coupling limit first.
The effective theory in low-energy limit is described by the zero modes at the edge since the bulk is gapped.
By the projection operator $P_e$ onto the edge states, the semi-infinite carbon nanotube is thus mapped into the finite $N_e$-cluster problem.
Since all edge states are pinned at zero energy, the hopping Hamiltonian after projection vanishes.
Only the interaction survives,
$H_{e} = U \sum_x P_e[n_{x\uparrow} n_{x\downarrow}]P_e \geq 0$. 
Consider the fully polarized state with all edge states in the spin-up states.
It is clear that the fully polarized state is an eigenstate of $H_e$ with zero eigenvalue.
Since the projected Hamiltonian is positive definite, it is the desired ground state with non-vanishing edge moment,
\begin{eqnarray}
M = N_e \mu_{B} = \left[ \frac{N-M+1}{3} \right] \mu_{B}.
\label{moment}
\end{eqnarray}
The weak-coupling result can be directly generalized to arbitrary interaction strength by the Lieb's theorem~\cite{Lieb89} which ensure the ground state is unique up to the trivial $(2S+1)$-fold degeneracy.

The metallic SWNTs are complicated by the pair of one dimensional conducting channels from the Dirac cones.
We can follow similar procedure to map out the gapped states in the bulk and derive the effective theory for the remaining edge states and the 1D conducting channels.
Again, the on-site interaction after projection prefers fully polarized edge states.
However, due to the interaction between the edge states and the conducting channels, the magnitude of the total spin can not be determined.
One may worry that Kondo-like screening from the itinerant carriers can further suppress the edge moment, or even make it vanish.
Evaluating the exchange integral between the edge and itinerant states, we found the exchange coupling is {\em ferromagnetic}, which becomes irrelevant under renormalization group transformation~\cite{Hewson97}.
Therefore, it is natural to expect that the edge moment will survive but its magnitude is beyond the reach of the weak-coupling approach.

So far, we show that the quantized magnetic moment in a semiconducting SWNT only depends on its chiral vector.
However, it is not clear whether this beautiful quantization is realistic or an artifact of the simple Hubbard model.
Therefore, we also carry out first-principles calculations within the local spin-density approximation, that takes account of the realistic band structure and the Couloumb interaction.
The self-consistent band structure calculations under lattice optimization were performed using the full-potential projected augmented wave method ~\cite{Kresse99} as implemented in the VASP package ~\cite{Kresse96}.
As an illustrating example, we compute the spin density profile in the $(10,2)$ chiral SWNT as shown in Fig. 1.
The semi-inifite nanotube was simulated by a finite segment, with one of the edges being the desired shape while with the other being the armchair (not shown in the figure) without additional edge states.
We have checked carefully that our results do not depend on the length of the segment and no extra moment is spotted near the armchair edge either.
When summing all spin densities on the $B$-sites, the total moment turns out to be $M=3 \mu_B$, as predicted in Eq.~\ref{moment}.
Furthermore, the nodal structure of the spin density is manifest and serves as a direct evidence for the approximate SUSY in nanotubes even with the inclusion of realistic band structures.

\begin{figure}
\centering
\includegraphics[width=7cm]{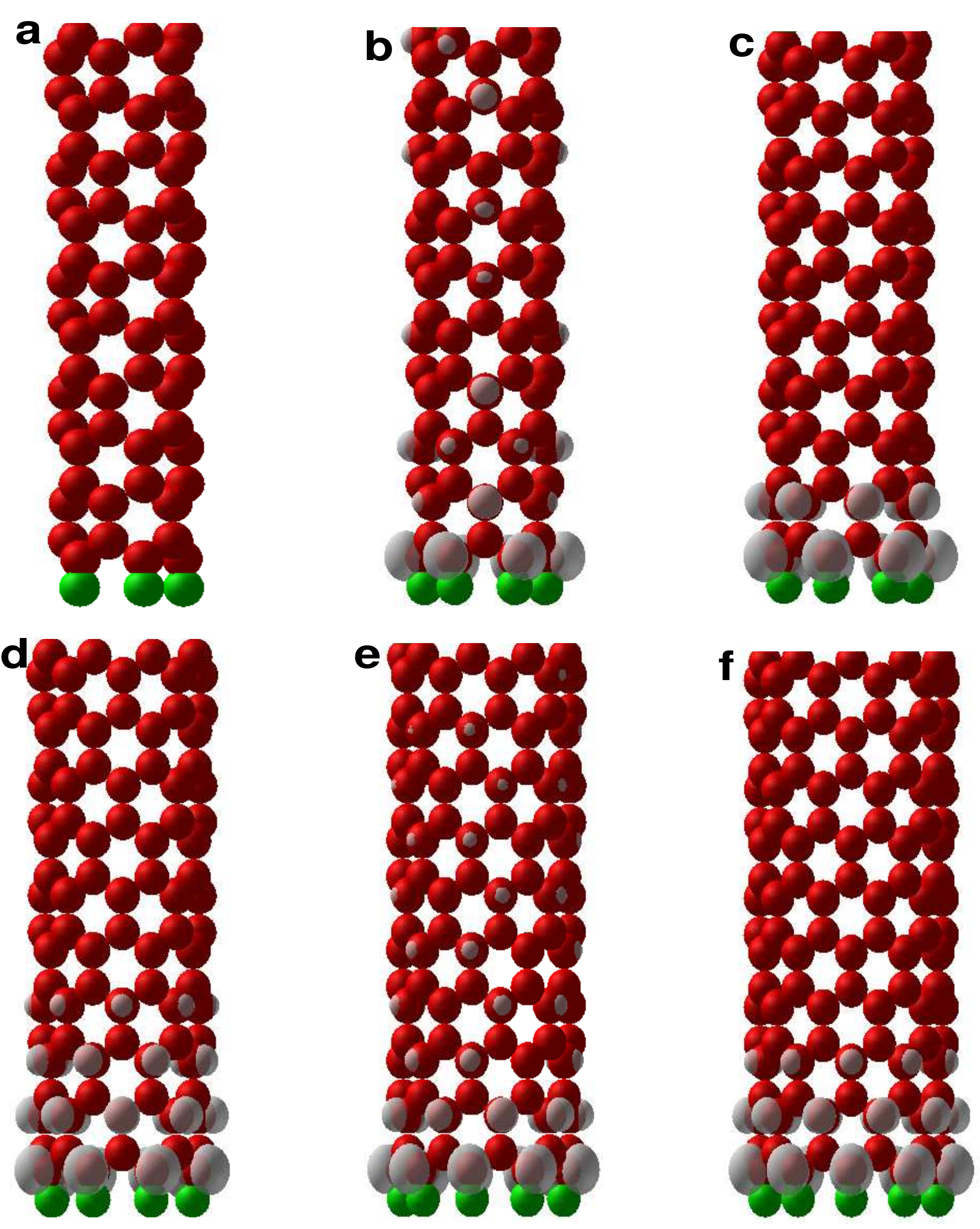}
\caption{\label{Fig3}
Edge moments in zigzag carbon nanotubes of various radii.
The spin density distributions for (a) $N=5$ (b)$N=6$ (c) $N=7$ (d) $N$=8 (e) $N$ =9 (f) $N$ =10 are shown in grey color while the red and green spheres denote C and H atoms respectively. Except for the small nanotubes (a) and (b), the magnetic moment is localized near the edge.}
\end{figure}

We also explore how the moment changes as the radius of the nanotube varies. Concentrating on the simplest zigzag SWNTs, the computed spin density profiles for various radii are shown in Fig. 3.
For the smallest radius $N=5$, no spin polarization is found.
For $N=6$, the moment starts to appear but is not localized to the edge.
For larger radii where the curvature effect is negligible, the edge moments are localized.
As long as the moment is localized to the edge, the magnetic moment, as shown in Fig. 4, follows the quantization rule in Eq.~(\ref{moment}) even for the metallic SWNTs!
For semiconducting ones $N=7,8,10$, the spin density profiles, shown in Fig. 3(c), (d), (f), decay rather quickly into the bulk and agrees with the predictions from the Hubbard model.
For the metallic one $N=9$ in Fig. 3(e), the spin density extends a bit further into the bulk due to the exchange coupling between the edge and the itinerant states but the magnetic moment remains quantized. The Lieb's theorem seems to provides a natural explanation for the quantized moment within the Hubbard model approximation. But, it remains an open question why the quantization is so robust even in a more realistic model,
as seen in the first-principles calculation.
In addition, our numerical results show that the hidden SUSY remains valid to high precision as the spin density exhibits the nodal structure. 

\begin{figure}
\centering
\includegraphics[width=6.5cm]{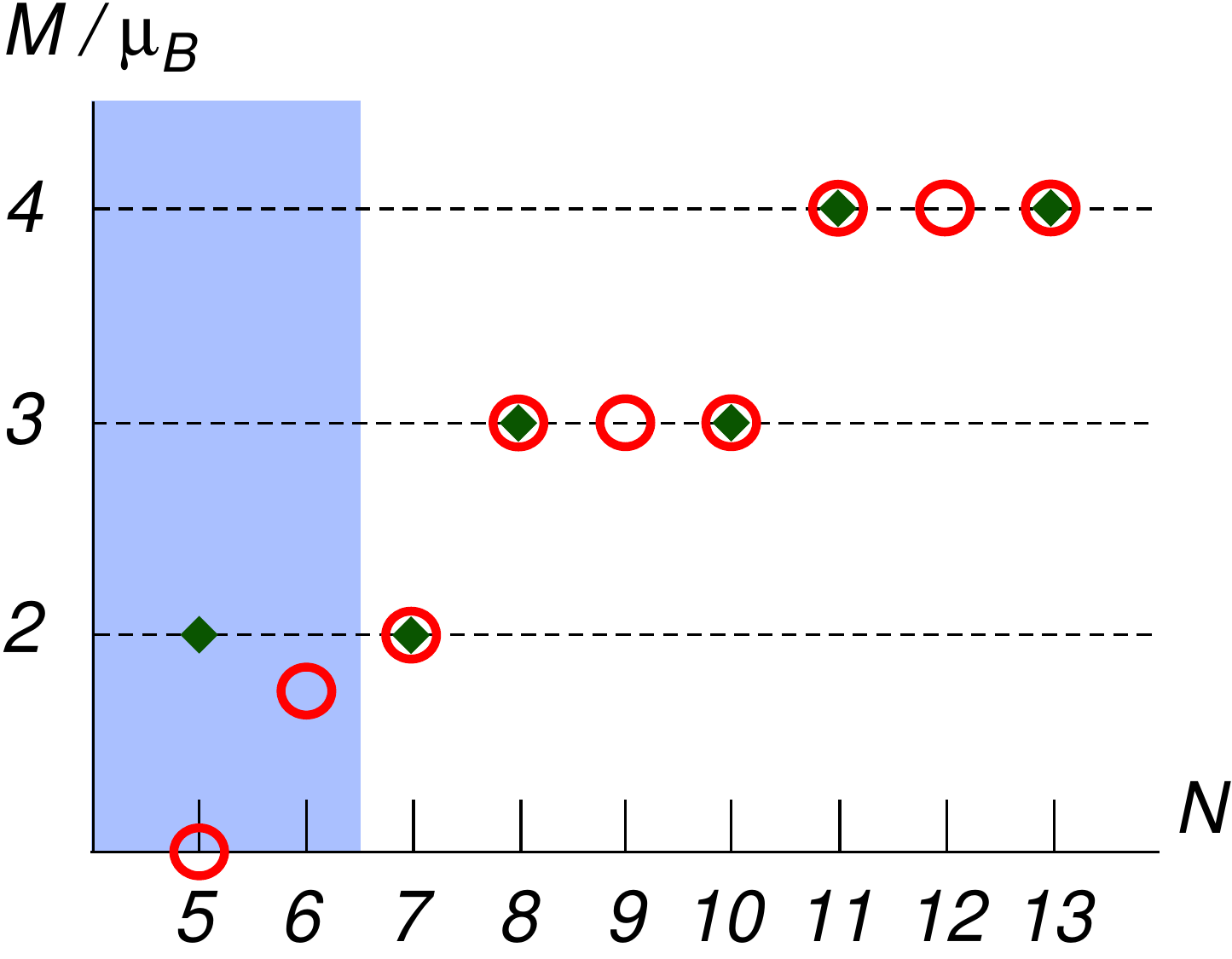}
\caption{\label{Fig4}
Magnitude of magnetic moment in zigzag SWNTs.
The green filled diamonds represent the predictions from analytic calculations for semiconducting nanotubes. The red circles mark the numerical results from the first-principle calculations. Both approaches give robust quantization of the edge moment except for small circumferences in the shaded blue regime. For $N \leq 6$, the curvature effect can not be ignored and ruins the edge moment.}
\end{figure}

In comparison with the nonmagnetic state, the energy gain of the spin quantized CNTs with N=7$\sim$13 range from 0.12 to 0.30 eV per cell, respectively, which correspond to about 20 meV per edge carbon for all the cases studied.
To understand the deviations at $N=5,6$, we studied a corresponding fictitious system of a flat graphene ribbon
with periodic boudary condition along the chiral vector,
by the same numerical method.
Although the system is topologically equivalent to the nanotube,
we find that the spin density is localized and the quantization of the
total spin is restored by the fully spin polarized flat edge bands.
Therefore, we conclude that the deviations found for nanotubes
with smaller radii arise from
curvature effects which also agree with previous studies~\cite{Hamada92}
on the band structure in the literature.
It is straightforward to generalize the numerical calculations for other
chiral SWNTs as well.

We have applied both analytic and numerical methods to establish the emergence of the quantized magnetic moment near the edge of a SWNT.
The edge moment can be detected (and perhaps manipulated) by spin-polarized scanning tunnelling microscopy~\cite{Pietzsch01} or by magnetic resonant force microscopy~\cite{Rugar04}.
Because of the robustness of the edge moments in the SWNTs, it provides an excellent method to fabricate nanomagnets.
Our findings demonstrate that the interplay between the electronic correlation and the edge morphology is of crucial importance at nanoscale.
The edge moment, solely depending on the topology of the nanotube, provides an excellent candidate for nanomagnet farbrication which adds a spin flavor to biological~\cite{Heller06} and chemical detections~\cite{Snow05} and also potential applications in spintronics~\cite{Ando00,Huertas-Hernando06,Nowack07,Kuemmeth08} for carbon nanotubes.
On top of the surprising phenomena for correlated physics in one dimension, carbon nanotubes can be used for biological and chemical detections at nanoscale as well.
The presence of the edge moment may enhance the performance with spin resolution.
In recent experiment, it was demonstrated that the spin and the orbital motions of electrons in carbon nanotubes are coupled strongly. Thus, when electron passing through the edge moment, the spin-orbit coupling provides another knob to control the transport properties and may provides another window for spintronics applications.

We acknowledge Leon Balents and Matthew Fisher for valuable discussions.
We acknowledge supports from the National Science Council in Taiwan through Grant No. NSC-97-2112-M-007-022-MY3.
The hospitality of KITP in Santa Barbara is also greatly appreciated.

\end{document}